\documentclass[a4paper,twocolumn,accepted=2026-03-03]{quantumarticle}
\pdfoutput=1  

\usepackage{graphicx}
\usepackage{amsmath,amssymb,amsfonts}
\usepackage[colorlinks, breaklinks=true]{hyperref}
\usepackage{color}
\usepackage[normalem]{ulem}

\usepackage[square,numbers,compress]{natbib}
\usepackage{xspace}

\renewcommand{\k}{{\bm k}}
\newcommand{\x}{{\bm x}}

\newcommand{\y}{{\bm y}}

\newcommand{\dbar}{\mathrm{d}\hspace*{-0.2em}\bar{}\hspace*{0.2em}}
\newcommand{\avg}[1]{\langle#1\rangle}
\newcommand{\CC}[0]{\mathbb{C}\xspace}
\newcommand{\RR}[0]{\mathbb{R}\xspace}

\newcommand{\dd}{\mathrm{d}\xspace}
\newcommand{\ket}[1]{| #1 \rangle}

\newcommand{\braket}[2]{\langle #1 | #2 \rangle}
\newcommand{\ketbra}[2]{| #1 \rangle\!\langle #2 |}
\newcommand{\matrixel}[3]{\langle #1 | #2 | #3 \rangle}

\usepackage{tikzit}

\tikzstyle{trapezium}=[fill=white, draw=black, shape=trapezium, trapezium stretches=true, trapezium right angle=-75, trapezium left angle=0, minimum width=1cm, minimum height=0.5cm, line width=0.025cm]
\tikzstyle{bigtrapezium}=[fill=white, draw=black, shape=trapezium, trapezium stretches=true, trapezium right angle=-75, trapezium left angle=0, minimum width=1.6cm, line width=0.025cm]
\tikzstyle{BIGtrapezium}=[fill=white, draw=black, shape=trapezium, trapezium stretches=true, trapezium right angle=-75, trapezium left angle=0, minimum width=1.8cm, line width=0.025cm]
\tikzstyle{daggerbigtrapezium}=[fill=white, draw=black, shape=trapezium, trapezium stretches=true, trapezium right angle=75, trapezium left angle=0, minimum width=1.6cm, line width=0.025cm]
\tikzstyle{state}=[fill=white, draw=black, shape=isosceles triangle, isosceles triangle stretches=false, inner sep=0, shape border rotate=-90, isosceles triangle apex angle=75, minimum width=0.7cm, line width=0.025cm, shape border uses incircle]
\tikzstyle{effect}=[fill=white, draw=black, shape=isosceles triangle, isosceles triangle stretches=false, inner sep=0, shape border rotate=90, isosceles triangle apex angle=75, minimum width=0.7cm, line width=0.025cm, shape border uses incircle]
\tikzstyle{discard}=[shape=tlground, fill=white, draw=black, rotate=180, scale=1.2]
\tikzstyle{bistate}=[fill=white, draw=black, shape=isosceles triangle, isosceles triangle stretches=false, inner sep=0, shape border rotate=-90, isosceles triangle apex angle=80, minimum width=1cm, line width=0.025cm, shape border uses incircle]
\tikzstyle{identity}=[shape=tlground, fill=white, draw=black, scale=1.2]
\tikzstyle{SAdiscard}=[shape=ground, fill=white, draw=black, rotate=180]
\tikzstyle{SAidentity}=[shape=ground, fill=white, draw=black]
\tikzstyle{square}=[fill=white, draw=black, shape=rectangle, minimum width=0.7cm, minimum height=0.7cm, line width=0.025cm, inner sep=0, rounded corners]
\tikzstyle{rectangle}=[fill=white, draw=black, shape=rectangle, minimum width=1cm, minimum height=0.7cm, line width=0.025cm, rounded corners]
\tikzstyle{longrectangle}=[fill=white, draw=black, shape=rectangle, minimum width=1.5cm, minimum height=0.7cm, line width=0.025cm, rounded corners]
\tikzstyle{looongrectangle}=[fill=white, draw=black, shape=rectangle, minimum width=3cm, minimum height=0.7cm, line width=0.025cm]

\tikzstyle{black}=[-, draw=black, line width=0.025cm]
\tikzstyle{da}=[-, dashed, gray]
\tikzstyle{arrow}=[->, draw=black, line width=0.025cm]
\tikzstyle{gray}=[-, draw=gray]
\tikzstyle{doublearrow}=[<->, line width=0.025cm, draw=black]

\input{basic.tikzdefs}

\usepackage[T1]{fontenc}

\usepackage{float} 
\usepackage{bm}

\usepackage[english]{babel}

\usepackage{ifthen}
\usepackage[normalem]{ulem}
\newcommand{\trackchanges}{true} 

\definecolor{darkgreen}{RGB}{0,128,42}
\definecolor{darkred}{RGB}{179,0,0}

\newcommand{\removed}[1]{%
\ifthenelse{\equal{\trackchanges}{true}}%
{{\small{\sout{#1}}}}%
{\xspace}%
}

\newcommand{\added}[1]{%
\ifthenelse{\equal{\trackchanges}{true}}%
{{\color{darkgreen}{#1}}}%
{#1}%
}%

\usepackage{xcolor}
\usepackage{xspace}
\usepackage{ifthen}

\newcommand{\showcomments}{false}

\newcommand{\andrea}[1]%
{\ifthenelse{\equal{\showcomments}{true}}%
{{\color{orange}{\small \textbf{Andrea says:} #1}}}{\xspace}}%

\newcommand{\marios}[1]%
{\ifthenelse{\equal{\showcomments}{true}}%
{{\color{blue}{\small \textbf{Marios:} #1}}}{\xspace}}%

\newcommand{\richard}[1]%
{\ifthenelse{\equal{\showcomments}{true}}%
{{\color{purple}{\small \textbf{Richard says:} #1}}}{\xspace}}%

\newcommand{\caslav}[1]%
{\ifthenelse{\equal{\showcomments}{true}}%
{{\color{red}{\small \textbf{\v Caslav says:} #1}}}{\xspace}}%

\newcommand{\carlo}[1]%
{\ifthenelse{\equal{\showcomments}{true}}%
{{\color{purple}{\small \textbf{Carlo:} #1}}}{\xspace}}%

\begin{document}

\title{Circuit locality from relativistic locality in scalar field mediated entanglement}

\author{Andrea {Di Biagio}}
\affiliation{Institute for Quantum Optics and Quantum Information (IQOQI) Vienna, Austrian Academy of Sciences, Boltzmanngasse 3, A-1090 Vienna, Austria}
\affiliation{Basic Research Community for Physics e.V., Mariannenstraße 89, Leipzig, Germany}

\author{Richard Howl}
\affiliation{Physics Department, Royal Holloway, University of London, Egham, Surrey, TW20 0EX, UK}
\affiliation{Quantum Group, Department of Computer Science, University of Oxford, Wolfson Building, Parks Road, Oxford, OX1 3QD, United Kingdom}

\author{\v Caslav Brukner}
\affiliation{Institute for Quantum Optics and Quantum Information (IQOQI) Vienna, Austrian Academy of Sciences, Boltzmanngasse 3, A-1090 Vienna, Austria}
\affiliation{Vienna Center for Quantum Science and Technology (VCQ), Faculty of Physics, University of Vienna, Boltzmanngasse 5, A-1090 Vienna, Austria}
\affiliation{Research Platform TURIS, University of Vienna, Vienna, Austria.}

\author{Carlo Rovelli}
\affiliation{Aix-Marseille University, Universit\'e de Toulon, CPT-CNRS, Marseille, France,}
\affiliation{Department of Philosophy and the Rotman Institute of Philosophy, Western University, London ON, Canada,}
\affiliation{ Perimeter Institute, 31 Caroline Street N, Waterloo ON, Canada}

\author{Marios Christodoulou}
\affiliation{Institute for Quantum Optics and Quantum Information (IQOQI) Vienna, Austrian Academy of Sciences, Boltzmanngasse 3, A-1090 Vienna, Austria}
\affiliation{Vienna Center for Quantum Science and Technology (VCQ), Faculty of Physics, University of Vienna, Boltzmanngasse 5, A-1090 Vienna, Austria}

\begin{abstract} 
\noindent Locality is a central notion in modern physics, but different disciplines understand it in different ways. Quantum field theory focuses on relativistic locality, based on spacetime regions, while quantum information theory focuses circuit locality, based on the notion of subsystems. Here, we investigate how spacetime and subsystem locality are related in the context of systems getting entangled while interacting via a scalar field. We show how, when the systems are put in a quantum-controlled superposition of localised states, relativistic locality (in the form of microcausality) gives rise to a specific kind of circuit. The relation between these forms of locality is relevant for understanding whether it is possible to formulate quantum field theory in quantum circuit language, and has bearing on the recent discussions on low-energy tests of quantum gravity.
\end{abstract}

\maketitle

\bigskip

\noindent The discovery that fundamental  interactions are mediated by fields has led to the modern idea of locality: there is no action at a distance. 
But action between what?

Relativistic locality says that no causal influence can propagate faster than the speed of light, and therefore there can be no causation between spacelike separated regions. In quantum field theory, this is ensured by \textit{microcausality}, also known as Einstein causality, the vanishing of the commutator between field operators at spacelike separations:
\begin{equation}
	[\hat\phi(x),\hat\phi(x')]=0
\end{equation}
if $x-x'$ is spacelike. This property, either shown to hold in physically relevant field theories or taken as an axiom in algebraic quantum field theory \cite{duncan2017conceptual}, allows for simultaneous measurements at different locations to be compatible and prevents superluminal signalling \cite{eberhard1989quantum}. Relativistic locality is a spatiotemporal notion of locality.

One can also formulate a notion of locality based on subsystems, where information can propagate from a system $A$ to system $B$ either via direct interaction between $A$ and $B$, or via a chain of successive interactions with some intermediary systems. We will call this notion \textit{subsystem locality}. In quantum theory, subsystem locality is articulated in terms of \textit{circuit locality}, where the evolution of several systems is modelled by a circuit, and the way the various gates are connected dictates the information flow between the subsystems \cite{lorenz2021causal}.

For example, if a unitary gate decomposes in the following circuit
\begin{equation}\label{subsystem_local_ex}
  \begin{tikzpicture}
	\begin{pgfonlayer}{nodelayer}
		\node [style=none] (1) at (-3.5, 2) {};
		\node [style=none] (2) at (-3.5, -2) {};
		\node [style=none] (3) at (-2.5, 2) {};
		\node [style=none] (4) at (-2.5, -2) {};
		\node [style=none] (5) at (-1.5, 2) {};
		\node [style=none] (6) at (-1.5, -2) {};
		\node [style=longrectangle] (11) at (-2.5, 0) {$\hat U$};
		\node [style=none] (14) at (1.5, 2.5) {};
		\node [style=none] (15) at (1.5, -2.5) {};
		\node [style=none] (16) at (2.5, 2.5) {};
		\node [style=none] (17) at (2.5, -2.5) {};
		\node [style=none] (18) at (3.5, 2.5) {};
		\node [style=none] (19) at (3.5, -2.5) {};
		\node [style=none] (27) at (0, 0) {$=$};
		\node [style=rectangle] (28) at (2, -1) {$~\hat U_1$};
		\node [style=rectangle] (29) at (3, 1) {$~\hat U_2$};
	\end{pgfonlayer}
	\begin{pgfonlayer}{edgelayer}
		\draw [style=black] (1.center) to (2.center);
		\draw [style=black] (3.center) to (4.center);
		\draw [style=black] (5.center) to (6.center);
		\draw [style=black] (14.center) to (15.center);
		\draw [style=black] (16.center) to (17.center);
		\draw [style=black] (18.center) to (19.center);
	\end{pgfonlayer}
\end{tikzpicture}

\end{equation}
one is assured that measurements on the system on the left can be affected by the middle system only and not by the system on the right. 
Circuit locality for circuits composed of unitary (or, more generally, completely positive trace-preserving maps) is a property of quantum mechanics. A version of circuit locality also holds in post-quantum theories, such as operational probabilistic theories \cite{dariano2017quantum} and generalised probabilistic theories \cite{barrett2007information}.

What is the precise relation between these two ideas of locality, one native of quantum field theory and one native of quantum information theory? Circuit locality in quantum theory is sometimes taken to reflect, encode, or ensure relativistic locality; see \cite[Section 6.3]{coecke2017picturing} and \cite{deutsch2000information}. The idea is that if we can locate the various gates of a circuit in different spacetime regions and have connections between gates located in causally connected regions, then the mathematics of Hilbert space ensures relativistic locality via its enforcement of circuit locality. This would explain why it is not possible to superluminally signal using entangled systems. The motivation is also present in reverse: the very identification of subsystems with Hilbert space tensor factors is sometimes motivated via appeals to relativistic locality in works in the foundations of quantum theory \cite{raymond-robichaud2021localrealistic,barrett2007information,dariano2017quantum}. However, the relation is not quite clear, with the main formal obstruction in motivating circuit locality from relativistic locality being the well-known result that it is impossible to decompose the Hilbert space of a quantum field theory in a tensor product of spaces associated with different regions \cite{witten2022why}. Delineating the connection between these two notions of locality is an open problem \cite{vanderlugt2024causally,vilasini2024fundamental}.

This connection matters in the discussion on the recent idea of witnessing nonclassical behaviour of gravity via gravity-mediated entanglement: two spatially separated masses becoming entangled as a result of the gravitational interaction  \cite{bose2017spin,marletto2017gravitationallyinduced,krisnanda2017revealing,Christodoulou2018c,christodoulou2023locally,martin-martinez2023what,fragkos2023inference,huggett2022quantum}. The study of these proposed experiments has created a new interface between the quantum gravity and  quantum information research communities.
The following question arises. Suppose there are two systems, $A$ and $B$, each coupled to a relativistically local field $\phi$ but not to each other. Does it follow that the field \textit{mediates} the interaction between $A$ and $B$, in the information-theoretic sense that the evolution of the full system can be broken down into a product of unitaries each acting only on $A$ and $\phi$, or on $B$ and $\phi$, and never involving both $A$ and $B$?

The Suzuki-Trotter theorem says that
\begin{equation}
	  \hat H=\hat H_{\phi A}+\hat H_{\phi B},
\end{equation}
we have 
\begin{equation}
  e^{-i\hat Ht} = \lim_{N\rightarrow\infty} \left(e^{-i\hat H_{\phi A}t/N}e^{-i\hat H_{\phi B}t/N}\right)^N,
\end{equation}
and since $\hat H_{\phi A}$ acts only on $\phi$ and $A$, and $\hat H_{\phi B}$ only on $\phi$ and $B$, it follows that the evolution can be approximated to any desired precision by an evolution with the required circuit locality.
This result however teaches us nothing about the relation between the two notions of locality we are interested in: the expansion relies only on general considerations about linear operators; relativistic locality plays no role in this.

In the following, we will go beyond the Suzuki-Trotter expansion and ask under which physical conditions---and in particular under which spacetime localisation assumptions---relativistic locality itself enforces a specific \textit{finite} circuit structure, with gates acting over finite time intervals which depend on the spatial configuration of the system. The resulting circuit structure is therefore informative precisely because it would generically fail in non-relativistic field theories, where microcausality does not constrain the relevant commutators.

\begin{figure*}
	\centering
	\includegraphics[width=1.8\columnwidth]{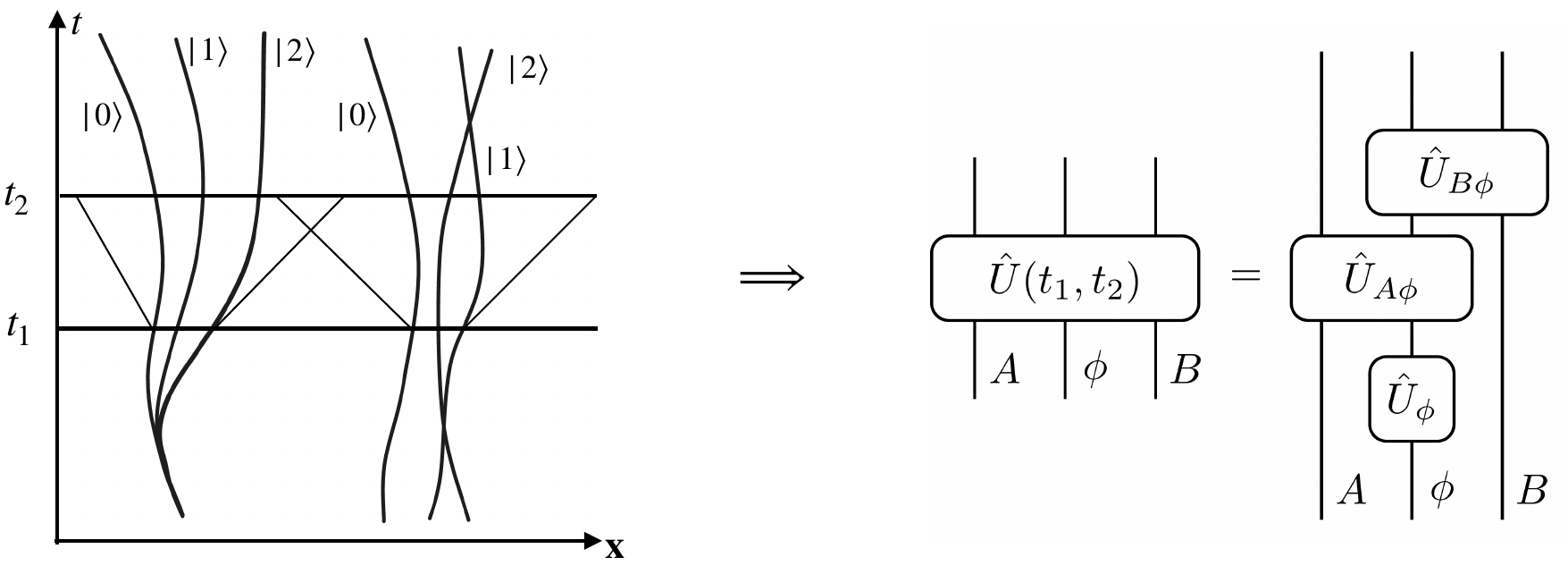}
	\caption{If the particles are in a quantum-controlled superposition of well-localised trajectories, the evolution of the system takes the desired field mediation shape in each time interval where all trajectories of one particle are spacelike separated from all trajectories of the other particle.}
	\label{fig:result}
\end{figure*}

In particular, we will show that if the systems $A$ and $B$ are kept spatially separated, each in a quantum controlled superposition of semiclassically localised states, microcausality allows to write the evolution of the three systems as
\begin{equation}
\label{eq:generalResult}
  \hat U(t_\mathrm{i},t_\mathrm{f}) = \prod_{n=1}^{N-1} \left(\hat U_{B\phi}^{(n)} \, \hat U_{A\phi}^{(n)} \hat U_\phi^{(n)} \right),
\end{equation}
where $\hat U_\phi$ acts only on the field, and $\hat U_{A\phi}^{(n)}$ and $\hat U_{B\phi}^{(n)}$ act only on the field and one other system, and where each value of $n$ represents the evolution over a \textit{finite time interval} $(t_n, t_{n+1})$ in which the particles are localised\footnote{More precisely: such that the components of the particles' wavefunction outside these spatially separated regions have negligible effect on the field.} in spacelike separated regions.
Restricting attention to any one such interval, our result can be written in diagrammatic notation as
\begin{equation}\label{bigUSLO}
    \begin{tikzpicture}
	\begin{pgfonlayer}{nodelayer}
		\node [style=longrectangle] (0) at (-2, 0) {$\hat U^{(n)}$};
		\node [style=none] (1) at (-3, 2) {};
		\node [style=none] (2) at (-3, -2) {};
		\node [style=none] (3) at (-2, 2) {};
		\node [style=none] (4) at (-2, -2) {};
		\node [style=none] (5) at (-1, 2) {};
		\node [style=none] (6) at (-1, -2) {};
		\node [style=none] (7) at (0, 0) {$=$};
		\node [style=rectangle] (9) at (1.5, 0) {$\hat U_{A\phi}^{(n)}$};
		\node [style=rectangle] (10) at (2.5, 2) {$\hat U_{B\phi}^{(n)}$};
		\node [style=none] (11) at (1, -3.5) {};
		\node [style=none] (12) at (2, -3.5) {};
		\node [style=none] (13) at (3, -3.5) {};
		\node [style=none] (14) at (1, 3.5) {};
		\node [style=none] (20) at (2, 3.5) {};
		\node [style=none] (25) at (3, 3.5) {};
		\node [style=square] (27) at (2, -2) {\makebox[0cm]{$\hat U_{\phi}^{\!(n)}$}};
	\end{pgfonlayer}
	\begin{pgfonlayer}{edgelayer}
		\draw [style=black] (1.center) to (2.center);
		\draw [style=black] (3.center) to (4.center);
		\draw [style=black] (5.center) to (6.center);
		\draw [style=black] (11.center) to (14.center);
		\draw [style=black] (12.center) to (20.center);
		\draw [style=black] (13.center) to (25.center);
	\end{pgfonlayer}
\end{tikzpicture}
~~~~~;
\end{equation}
see figure~\ref{fig:result}.
Additionally,
\begin{equation}\label{commutes}
	\hat U_{A\phi}^{(n)}\hat U_{B\phi}^{(n)} = \hat U_{B\phi}^{(n)}\hat U_{A\phi}^{(n)},
\end{equation}
showing that the disturbance one system may have on the field cannot propagate to the other system via the field in these time intervals. We will call a unitary satisfying \eqref{eq:generalResult} and \eqref{commutes} a \textit{field mediation}.

In contrast with the Suzuki-Trotter version of this decomposition, here the role of spacetime locality is evident: each term in the product in \eqref{eq:generalResult} evolves the system for a finite duration, given by the separation between the systems $A$ and $B$. Given \eqref{commutes}, we also see that, during this time interval, $A$ and $B$ cannot affect each other but only the field~\cite{lorenz2021causal}, as one would expect for spacelike-separated systems.

The result is easily generalised to the case of an arbitrary number of particles. The approximations we employ to obtain this result are appropriate to describe a number of physically relevant regimes: quantum computing platforms, such as ion traps; quantum optics experiments, such as Bell tests; and the regime in which low energy tests of quantum gravity may be performed.

In the following four sections we derive this result. After laying down some notation in Section~\ref{sec:setup}, we define the required approximations and derive their consequences in Section~\ref{sec:param_main}. In Section~\ref{sec:field}, we derive an expression for the evolution of a quantum field coupled to a classical source, incidentally improving on textbook results by taking into account  a phase which is often neglected. This phase plays a crucial role in our result, as we will see in Section~\ref{sec:main_result}, where we demonstrate how relativistic locality influences circuit locality in our system. In Section~\ref{sec:entanglement}, we show how this setup allows for scalar field mediated entanglement by deriving the explicit, relativistic expression for the entangling phases. In the conclusion, we discuss the limitations and possible generalisations of our result. Throughout the paper, we refer the reader to the appendix, where we report more detailed derivations of the results.

\section{Setup}
\label{sec:setup}

We consider a massive scalar field coupled to two quantum sources, coherently controlled by two control qudits ($d$-level systems with Hilbert space $\CC^d$).\footnote{We work in units such that $\hbar=c=1$. We denote four-vectors as $x$ and space vectors as $\bm x$. For brevity, we will sometimes suppress the explicit dependence on the interval $(t,t')$ for unitary evolution operators and other quantities. We will use hats to denote operators and distinguish them from c-numbers.} Coherent control with qudits is a typical setup in quantum information theory \cite{nielsen2010quantum}. The Hilbert space for the field is the usual Fock space $\mathcal F$ generated by the canonical creation-annihilation operators acting on the vacuum of the free Hamiltonian of the field. The Hilbert spaces for the particles are $A=L^2(\RR^3)\otimes \mathcal H_A$ and $B=L^2(\RR^3)\otimes \mathcal H_B$, where $L^2(\RR^3)$ represents centre of mass motion and $\mathcal H_A$ and $\mathcal H_B$ some internal degrees of freedom of the particles. 
The Hamiltonian of the system is taken to be
\begin{equation}\label{full_Hamiltonian}
    \hat H(t) = \hat H_A(t) + \hat H_B(t) + \hat H_0 + \hat H_\mathrm{int}.
\end{equation}
The Hamiltonian $\hat H_A(t)$ encodes the dynamics of $A$ as influenced by a (possibly time-dependent) driving by the first control qudit. It acts only on these two systems and is of the form 
\begin{equation}
\hat H_A(t) = \sum_{r=1}^d\ketbra rr\otimes \hat H^r_A(t),    
\end{equation}
with $\ket{r}$ denoting the computational basis of the first qudit and $\hat H^r_A(t)$ acting only on $A$; similar definitions hold for $\hat H_B(t)$. The free Hamiltonian of the field is given as usual by
\begin{equation}
	\hat H_0 = \int\frac{\dd^3k}{(2\pi)^3}\,\omega_k\hat a^\dagger_\k\hat a_\k,
\end{equation}
with ${\omega_k=\sqrt{m^2+k^2}}$.

The interaction term couples the particle positions to the field. It is given by
\begin{equation}\label{Hint}
	\hat H_\mathrm{int} = \int\dd^3x\,\hat\phi(\x)(\hat\rho_A(\x)+\hat\rho_B(\x))
\end{equation}
where $\hat\rho_A$ and $\hat\rho_B$ are the charge density operators\footnote{This is the analog of electric charge density or mass density for the scalar field. For pointlike particles we would have ${\hat \rho_A(\x) = \ketbra \x\x_A}$, ie, ${\hat H_\mathrm{int}=\hat\phi(\hat\x_A)+ \hat\phi(\hat\x_B)}$. For something with spatial extent we would have ${\rho_A(\x)=\int\dd^3x_A\sigma_A(\x-\x_A)\ketbra{\x_A}{\x_A}_A}$ with $\sigma_A(\x)$ a real-valued function of compact support that represents the charge distribution conditional on the particle being localised at the origin.} for the particles.

\section{Parametric approximations}
\label{sec:param_main}

We now assume that the back reaction of the particles on the qudits, on the one hand, and of the field on the particles on the other, can be safely ignored. This kind of approximation is sometimes known as \textit{parametric} and is often used in quantum optics \cite{breuer2002theory,scully1997quantum}, and is also part of the Born-Oppenheimer approximation.

Concretely, we first assume that, at all times $t$, the state of the system is of the form
\begin{equation}\label{main_first_param_assumption}
\ket{\Psi(t)}=\frac1d\sum_{r,s\in\{1,\dots,d\}}\ket{rs}\ket{\Psi^{rs}(t)},
\end{equation}
where $\ket{rs}$ are the states in the computational basis of the qudits and $\ket{\Psi^{rs}(t)}$ are normalised states of the particles and field. 
Since the states $\ket{rs}$ are orthonormal and $\hat H(t)$ is diagonal in this basis, we have 
\begin{equation}\label{main_first_param_result}
   \frac{\dd}{\dd t} \ket{\Psi^{rs}(t)} = -i\hat H^{rs}(t)\ket{\Psi^{rs}(t)},
\end{equation}
with $\hat H^{rs}(t)=\matrixel{rs}{\hat H(t)}{rs}$.

To solve this equation, we further assume that, for all times $t$,
\begin{equation}\label{main_second_param_assumption1}
    \ket{\Psi^{rs}(t)} = \ket{\psi^r_A(t)}\ket{\psi^s_B(t)}\ket{\psi_\phi^{rs}(t)},
\end{equation}  
where $\ket{\psi^r_A(t)}, \ket{\psi^s_B(t)},$ and $\ket{\psi_\phi^{rs}(t)}$ are states for the particles $A$ and $B$, and the field, respectively.  
In physical terms, this approximation amounts to disregarding any entanglement between the field and a particle caused by the quantum spread of the particle in each individual $\ket{rs}$ branch.  In other words, we assume that the particles are put in a quantum-controlled superposition of pointer states, in the sense of Zurek \cite{zurek2003decoherence}. Note that this does not prevent the generation of entanglement in the full state \eqref{main_first_param_assumption} or the detection of entanglement between different subsystems; we will return to this point in section~\ref{sec:entanglement}. Similarly, we assume that
\begin{equation}\label{main_second_param_assumption2}
    \frac{\dd}{\dd t}\ket{\psi^r_A(t)}=-i\hat H^r_A(t)\ket{\psi^r_A(t)},
\end{equation}
with $\hat H^r_A(t)=\matrixel{r}{\hat H_A(t)}{r}$ and the same for $B$. It then follows that the field states $\ket{\psi_\phi^{rs}(t)}$ also satisfy a Schr\"odinger equation
\begin{equation}\label{main_second_param_result}
    \frac{\dd}{\dd t}\ket{\phi^{rs}(t)}=-i\hat H^{rs}(t)\ket{\psi_\phi^{rs}(t)},
\end{equation}
with $\hat H^{rs}(t) = \hat H_0+\hat H^{rs}_\mathrm{int}(t)$ and
\begin{equation}\label{Hrs_int}
  \hat H^{rs}_\mathrm{int}(t)=\matrixel{\psi^r_A(t)\psi^s_B(t)}{\hat H_\mathrm{int}}{\psi^r_A(t)\psi^s_B(t)}. 
\end{equation}
Thus, the unitary evolution will be of the form
\begin{equation}
  \hat U = 
        \sum_{rs}\ketbra{rs}{rs}\otimes
         \hat U_B^s \, \hat U_A^r \,
         \hat U_\phi^{rs},
\end{equation}
where $\hat U_A^r$ and $\hat U_B^s$ solve the Schr\"odinger equation \eqref{main_second_param_assumption2} and the analogous one for $B$, and $\hat U_\phi^{rs}$ solves \eqref{main_second_param_result}. 

Note that this is still not in the desired form, but it would be if we also had
\begin{equation}\label{wishSL}
   \hat U_\phi^{rs} = \hat U_\phi^{s} \,  \hat U_\phi^{r},
\end{equation}
for some unitaries $\hat U_\phi^{s}, \hat U_\phi^{r}$. Indeed, then we would have
\begin{equation}
  \hat U = \underset{\hat U_{\phi B}}{\underbrace{\left(\sum_s \ketbra ss  \hat U_B^s\hat U_\phi^s\right)}}\circ\underset{\hat U_{\phi A}}{\underbrace{\left(\sum_r \ketbra rr  \hat U_A^r\hat U_\phi^r\right)}}\,.
\end{equation}
 We therefore need to investigate the properties of $\hat U_\phi^{rs}$.

\section{Evolution of the field states} 
\label{sec:field}

In each $\ket{rs}$ component, the field state evolves according to $\hat H^{rs}$. Putting together  \eqref{Hrs_int} and \eqref{Hint} we see that
\begin{equation}
	\hat H^{rs}_\mathrm{int}(t) = \int\dd^3x \,\hat \phi(\x)\rho^{rs}(t,\x),
\end{equation}
where
\begin{equation}
	\rho^{rs}(t,\x)=\matrixel{\psi^r_A(t)\psi^s_B(t)}{\hat \rho(\x)}{\psi^r_A(t)\psi^s_B(t)},
\end{equation}
with $\hat \rho = \hat \rho_A+\hat \rho_B$.
Thus, $\hat H^{rs}$ describes the evolution of a quantum field coupled to a time-dependent classical source; this can be solved in closed form using standard methods. We will summarise the results in this section and detail the procedure in appendix~\ref{app:quantum-classical}.

The propagator $\hat U^{rs}_\phi(t,t')$ that evolves a field state $\ket{\phi^{rs}(t)}$ to $\ket{\phi^{rs}(t')}$ according to \eqref{main_second_param_result} can be written as
\begin{equation}\label{U_phi}
	\hat U^{rs}_\phi(t,t') = e^{-i\hat H_0t'}\hat U^{rs}_I(t,t')e^{i\hat H_0t},
\end{equation}
where $\hat U_I^{rs}$ denotes the interaction picture propagator. It is the solution to
\begin{equation}\label{int_phi_main}
  \frac{\partial }{\partial t'}\hat U_I^{rs}(t,t') = -i\hat H_I^{rs}(t')\hat U^{rs}_I(t,t'),
\end{equation}
where 
\begin{equation}
  \hat H_I^{rs}(t) 
  =\int\dd^3x\, \rho^{rs}(t,\x) \hat \phi_I(t,\x),
\end{equation}
with $\hat\phi_I(t,\x)$ the interaction picture field operators. Instead of expressing $\hat U^{rs}_I(t,t')$ as Dyson series time-ordered exponential, here it is advantageous to use the Magnus expansion 
\begin{equation}
  \hat U_I^{rs}(t,t')=\exp\left(\sum_{n=1}^\infty\hat\Omega^{rs}_n(t,t')\right),
\end{equation}
where the expression for the infinite sequence of operators $\hat\Omega^{rs}_n(t,t')$ can be found in \cite{blanes2009magnus}.
In our case, thanks to the fact that both $\rho^{rs}$ and $[\hat \phi_I(t,\x),\hat \phi(t',\x')]$ are c-numbers, we have that
$[\hat H_I(t),\hat H_I(t')]$ is a c-number and one can obtain a relatively simple equation for the interaction picture propagators:
\begin{equation}\label{app:UI-magnus}
\hat U_I^{rs}(t,t')=e^{\Omega^{rs}_2(t,t')}e^{\hat\Omega^{rs}_1(t,t')},
\end{equation}
with $\hat \Omega^{rs}_1$ and $\Omega^{rs}_2$ given in \eqref{Omega1_def} and  \eqref{Omega2-def}, respectively.
 
In appendix \ref{sec:physical-interpretation}, we show that the Schr\"odinger picture propagators can then be expressed in a physically intuitive way, namely,
\begin{equation}
    \hat U^{rs}_\phi(t,t') = e^{\Omega_2^{rs}(t'-t)}\hat{\mathcal D}[\rho^{rs}]e^{-i\hat H_0\cdot(t'-t)},
\end{equation}
where $e^{-i\hat H_0\cdot(t'-t)}$ is a free propagation of the field, followed by a displacement operator $\hat{\mathcal D}[\rho^{rs}]$ that takes into account the effect of the particles on the field between $t$ and $t'$, the whole thing multiplied by a phase factor $e^{\Omega_2^{rs}(t'-t)}$ that does not depend on the state of the field.

This is the well-known result that classical sources create coherent states of the field, derived in standard references such as \cite{cohen-tannoudji1989photons,gerry2005introductory}.  However, these standard treatments often omit the phase factor $e^{\Omega^{rs}_2}$. This is because, so long as we are only concerned with the effect of classical sources on the field, this is a global phase  which has no observational consequences. Here---because we want to be able to deal with quantum superpositions of sources---this phase must be considered carefully. Additionally, this phase is necessary to ensure the correct groupoid property of the unitary operators,
\begin{equation}\label{main:group-property}
  {{\hat U^{rs}_\phi(t,t'')=\hat U^{rs}_\phi(t',t'')}}\hat U^{rs}_\phi(t,t')
\end{equation}
 for $t\leq t'\leq t''$, whihc will allow us to break down the evolution over several time intervals and study its causal properties.

\section{Microcausality and circuit locality}
\label{sec:main_result}

We are now in a position to put everything together and see the connection between relativistic causality and circuit locality. 

Recall that ${\hat \rho = \hat \rho_A+\hat \rho_B}$, from which follows that ${\rho^{rs}=\rho_A^r+\rho_B^s}$.
We can then use the Baker-Campbell-Hausdorff formula to separate the contribution of each particle to the evolution of the field, obtaining (we refer the reader to Appendix~\ref{sec:two_classical_source} for more details)
\begin{equation}
\label{eq:almostSL}
      \hat U^{rs}_\phi(t,t') = e^{\Omega^{rs}(t,t')}\hat U^{r}_\phi(t,t') \, \hat U^{s}_\phi(t,t') \, e^{-i\hat H_0\cdot(t'-t)},
\end{equation}
where $\hat U^r_\phi=e^{\Omega_2[\rho^r_A]}\hat{\mathcal D}[\rho_A^r]$ 
computes the effect of the source $A$ alone, and similarly for $\hat U^s_B$. This operator is almost in the desired form \eqref{wishSL}, the only obstruction being the pure phase $e^{\Omega^{rs}}$, given by
\begin{widetext}
\begin{equation}
\begin{aligned}
   \Omega^{rs}(t,t') =
    &\iint_{t}^{t'}\dd \tau \dd \tau' \iint\dd^3 x\dd^3x'\, 
    \rho_A^r(\tau,\x)\rho_B^s(\tau',\x')
\, [\hat\phi_I(\tau,\x),\hat\phi_I(\tau',\x',)] \\
    &+ \int_{t}^{t'}\dd \tau \int_{t}^{\tau}\dd \tau'\iint\dd^3 x\dd^3x'\, \big(\rho_A^r(\tau,\x)\rho_B^s(\tau',\x')+\rho_A^r(\tau',\x')\rho_B^s(\tau,\x)\big)
    \, [\hat\phi_I(\tau,\x),\hat\phi_I(\tau',\x',)],
\end{aligned}
\end{equation}
\end{widetext}
with ${\rho^r_A(t,\x) = \matrixel{\psi^r_A(t)}{\hat \rho_A(\x)}{\psi^r_A(t)}}$, and similarly for $\rho^s_B$.
The first term arises from the Baker-Campbell-Hausdorff formula when splitting the displacement operator, the second term is due to the non-linearity of $\Omega_2^{rs}$; see \eqref{Omega2-def}.

It is this phase factor that can prevent the desired factorisation \eqref{wishSL} of $\hat U^{rs}$, and therefore the circuit locality of the full evolution hinges on the properties of this phase.
Since the commutator of interaction picture fields appears explicitly here, this is where microcausality comes into play.

Assume that the supports of $\rho^r_A$ and $\rho^s_B$ are spacelike separated (for each $rs$ pair) during the interval $[t,t']$. Products of the form $\rho_A^r(\tau,\x)\rho_B^s(\tau',\x')$ will then only be non-zero for $(\tau-\tau',\x-\x')$ spacelike, precisely when the commutators $[\hat\phi_I(\tau,\x),\hat\phi_I(\tau',\x')]$ vanish due to microcausality. It follows that ${\Omega^{rs}=0}.$ Then $\hat U^{rs}$ factorises as in \eqref{wishSL} and that the total evolution $\hat U$ is a field mediation in this time interval.

If the supports are not spacelike separated for the whole interval $[t,t']$, they may still be spacelike separated in sub-intervals $(t_n,t_{n+1})$ of finite length, such that $\min_n(t_{n+1}-t_n)>0$. The result would still hold because one can decompose the unitary evolution into many finite intervals in which the particles are not in causal contact using the group property \eqref{main:group-property}.

Let $d^{rs}(t)$ be the distance between the supports of $\rho_A^r$ and $\rho_B^s$ at time $t$, and define the distance of closest approach as $d_\mathrm{min}=\inf_t\min_{rs} d^{rs}(t)$. Within our approximations, then, the evolution falls into field mediation form if $d_\mathrm{min}>0$.

\section{Scalar field mediated entanglement}
\label{sec:entanglement}

We now show that this setup is capable of reproducing the scalar analog of the gravity-mediated entanglement as originally proposed in \cite{bose2017spin}, where the control qudits are embedded in the particles, and used to put them in spatial superpositions.

Suppose that for $t<t_\mathrm{i}$ the two particles are localised and at rest and the field is at equilibrium, so that the state is given by
\begin{equation}
	\ket{\Psi_\mathrm{i}}=\frac1d\sum_{rs}\ket{rs}\ket{\psi_{A,\mathrm i}}\ket{\psi_{B,\mathrm i}}\ket{\phi_\mathrm{i}},
\end{equation}
where $\ket{\phi_\mathrm{i}}$ is the coherent state of the field peaked around the static classical configuration $\phi_\mathrm{i}$ sourced by the particles at their initial locations.
Then between $t_\mathrm{i}$ and $t_\mathrm{f}$, the driving Hamiltonians $\hat H_A(t)$ and $\hat H_B(t)$ are such that they put the particles in a superposition of semiclassical localised states moving along spatially separated trajectories; see figure~\ref{fig:result}. The state of the system for $t\in[t_\mathrm{i},t_\mathrm{f}]$ is
\begin{equation}
	\ket{\Psi(t)}=\frac1d\sum_{rs}\ket{rs}\ket{\psi_{A}^r(t)}\ket{\psi_{B}^s(t)}\ket{\psi_\phi^{rs}(t)},
\end{equation}
where each subsystem is entangled with the rest.
The superposition of the particles is undone by $t_\mathrm{f}$ and the final state will be
\begin{equation}
    \ket{\Psi_\mathrm{i}}= \frac1d\sum_{rs}e^{i\theta^{rs}}\ket{rs}\ket{\psi_{A,\mathrm f}}\ket{\psi_{B,\mathrm f}}\ket{\phi_\mathrm{f}^{rs}},
\end{equation}
where $\ket{\phi_\mathrm{f}^{rs}}$ is the coherent field peaked around the classical configuration $\phi_\mathrm{f}^{rs}$ obtained by solving the Klein-Gordon equation with source $\rho^{rs}$ and initial condition $\phi_\mathrm{i}$. If we additionally assume that the experiment was done slowly enough such that the $\ket{\psi^{rs}_\mathrm{f}}$ have quite large overlap with each other, the control qudits are effectively disentangled from the rest of the system and their state is well approximated by the state
\begin{equation}
	\frac1d\sum_{rs} e^{i\theta^{rs}}\ket{rs},
\end{equation}
whose entanglement can be detected by standard methods \cite{bose2017spin,polino2024photonic}.
The phases $\theta^{rs}$ responsible for the entanglement have a neat physical interpretation. They are given by
\begin{equation}
	\theta^{rs} = -\frac12\int\dd^4x\,\rho^{rs}(x)\phi^{rs}(x),
\end{equation}
where $\phi^{rs}(t,\x)$ is the solution to the Klein-Gordon equation with initial conditions $\phi_\mathrm{i}$ and source $\rho^{rs}$. Since the Klein-Gordon equation is hyperbolic, the effect of each particle on $\phi^{rs}$ propagates causally, so that the phases can only entangle the two qudits only if there is enough time for light to travel from the region of one particle to that of the other, that is only if $t_\mathrm{f}-t_\mathrm{i}>d_\mathrm{min}$. This relativistic locality property is in complete alignment with that of the gravitational version of the experiment first derived in \cite{christodoulou2023locally}. In fact, the phases $\theta^{rs}$ can be interpreted as the on-shell values of the field sector action
\begin{equation}
	S_\phi^{rs} = \int\dd^4x\left(\frac12\phi(\square-m^2)\phi - \rho^{rs}\phi\right)
\end{equation}
corresponding to the Hamiltonian $\hat H^{rs}_\phi$, as already remarked in \cite{christodoulou2023locally}.

\section{Discussion}

We considered the evolution of a massive scalar field coupled to two quantum systems with positional and internal degrees of freedom, and we took these particles to be coherently controlled by a control qudit. Within this setting, we showed that the field is \emph{mediating} an interaction between the particles, not only in the sense of no direct particle-particle coupling in the Hamiltonian, but also in the precise sense provided by the field mediation quantum circuit.

We emphasize the fact that we have taken the assumption that a particle's effect on the field can be modelled, in each $\ket{rs}$ branch, by a source with compact support. In quantum mechanics, however, the wavefunction of a particle will not have bounded support for any finite amount of time,  unless it is confined by an infinite potential well (which itself needs to be perfectly localised). This implies that in the general situation the support of $\psi^r_A$ and $\psi^s_B$ will be overlapping, as will those of $\rho^r_A$ and $\rho^s_B$, and the evolution \textit{will not} take the field mediation shape. 
From the perspective of the quantum field theoretical aspects of nature, the notions of local particles and qudits, and their circuit local evolution are only approximations. Nevertheless, in concrete situations, we know from experience this approximation is appropriate: for all practical purposes, the location of a particle, and its sourcing of a field, can be well approximated as being contained in a compact region.

Let us  discuss the limitations of our computation.
The parametric approximation implies that we neglect the back-reaction of the field on the particles, and allowed us to do a ``branch-by-branch'' analysis. It also assumes that, in each $rs$ branch, the particles and the field do not get entangled, which is a good approximation only if the states $\ket{\psi^r_A(t)}$ and $\ket{\psi^s_B(t)}$ are quite localised at all times. These assumptions are commonly used to model superpositions of semi-classical scenarios, and yield many interesting phenomena; see 
\cite{bose2017spin,marletto2017gravitationallyinduced,mari2016experimentsa,belenchia2018quantum,danielson2022blacka,danielson2023killing} for a tiny sample.
Removing this approximation makes the analysis considerably more involved.
For example, when computing directly evolution due to the full Hamiltonian \eqref{full_Hamiltonian} using the Magnus expansion for the interaction picture, the non-trivial commutators of $\hat \rho$ prevent the expansion from terminating. The result would still hold perturbatively, at low enough order, but we leave a complete analysis to future work.

Another limitation to the physical significance of our model is the fact that we have studied a massive scalar field, while the fundamental interactions in physics are via massless gauge fields. One issue that will arise in the massless theory is that the displacement operators $\hat{\mathcal D}$ will no longer be well-defined unitary operators for all values of the source $\rho$. This is the well-known problem of infrared divergences in quantum field theory \cite{prabhu2022infrared,peskin2019introduction}. There are two ways to deal with this. One is to impose an infrared cutoff to the field, in which case the result holds immediately. Alternatively, one would have to embrace the fact that the algebra of observables will not be representable by operators on a single Hilbert space. We expect it should be possible to derive a similar result also in this setting, provided a notion of subsystem that does not rely explicitly on Hilbert spaces.

Since we worked with a scalar field, we did not have to deal with gauge constraints. In symmetry-reduced quantisation, the imposition of constraints can imply the presence of terms in the Hamiltonian that couple charged particles directly. For example, when quantising Maxwell's theory in the Coulomb gauge, we would have the Coulomb term $\propto 1/|\hat \x_A - \hat \x_B|$ spoiling mediation from the get go \cite{anastopoulos2018comment,anastopoulos2021gravitational, fragkos2023inference}.
This is similar to what happens in the classical theory: the Coulomb gauge spoils the manifest Lorentz locality of electromagnetism in favour of computational efficiency. However, it is only `manifest' Lorentz locality that is lost, all the physics remains unchanged and, in particular, relativistic causality still holds. Does a similar thing happen with subsystem locality in the quantum theory?

Dirac quantisation of gauge systems may be  more suitable for keeping subsystem locality explicit. For example, in Gupta-Bleuler quantisation for the electromagnetic field \cite{gupta1950theory,bleuler1950neue,cohen-tannoudji1989photons}, we have four components each coupled to the sources with no interaction terms coupling the sources directly. Thus, formally, we have four massless scalar fields coupled to the sources and the dynamics resembles that studied in the main text. The gauge redundancy is removed instead by imposing a constraint on the physical states, a constraint that is maintained by the equations of motion. Perhaps a similar result can be derived in this setting.

Finally, let us briefly comment on the significance of this result on the discussion surrounding the low energy experimental tests of quantum gravity. The generation of entanglement between two massive quantum systems via gravity has been argued to provide a theory-independent witness of the nonclassicality of the gravitational field \cite{bose2017spin,marletto2017gravitationallyinduced,krisnanda2017revealing,fragkos2023inference,huggett2022quantum}. This argument invokes theorems from quantum information theory and their post-quantum generalisations, which we may call `LOCC theorems'. These theorems state that a classical system cannot mediate the creation of entanglement between two systems. These no-go theorems are often presented as the dilemma: if gravity can create entanglement, then it is either non-local or non-classical.

The notion of locality at stake in these theorems is the quantum information notion of mediation, not that of relativistic locality~\cite{fragkos2023inference,dibiagio2025simple}. If no field theory obeyed the kind of locality in the LOCC theorems, the theorems would not have a strong bite, since a no-go theorem's potency lies in the extent to which its conflicting assumptions are considered desirable properties of physical theories.
Granted the above limitations, our result shows that the particular form of the locality assumption in the LOCC theorems is indeed a property of a relativistic quantum field theory in the appropriate regime, thus strengthening the relevance of these theorems to the low-energy test of quantum gravity debate. The next step would be to prove a similar result in linearised quantum gravity, at least in some appropriate gauge.

We expect this work to spark further investigations on the above open questions in order to generalise the result given here to the degree possible. In general, a relativistic quantum field theory will not give rise to circuits in with unitary gates embedded in regions of spacetime. At a minimum, certain localisation conditions have to be satisfied by the states of such a theory.
The end result of the effort started here would hopefully be a clear demarcation of the approximations and assumptions needed in order for physically relevant quantum field theories to be cast in a quantum information---quantum circuit---language. \\

\bigskip

\begin{acknowledgements}
We thank Davide Poderini, Gautam Satishchandran, Daine Danielson, Maria Papageorgiou, Borivoje Daki\'c, Markus Aspelmeyer, Ofek Bengyat, as well as Sougato Bose and his group, for enlightening discussions at various stages of of this work.

We acknowledge support of the ID\#~61466 grant from the John Templeton Foundation, as part of the ``Quantum Information Structure of Spacetime (QISS)'' project (\hyperlink{http://www.qiss.fr}{qiss.fr}), as well as the ID\#~63683 from the John Templeton Foundation as part of WOST, WithOut SpaceTime project (\href{https://withoutspacetime.org}{withoutspacetime.org}).  \v{C}.B. acknowledges support from the research platform TURIS. R.H.  acknowledges  support of grant ID\#~62420 from the John Templeton Foundation. The opinions expressed in this work are those of the author(s) and do not necessarily reflect the views of the John Templeton Foundation.
\end{acknowledgements}

\bigskip
\onecolumngrid

\appendix

\pagebreak

\section*{Appendix}

We work in units $\hbar=c=1$, denote 3-vectors in bold font, and use the shorthand ${\dbar^3k = \dd^3k/(2\pi)^3}$. The metric signature is ${(-,+,+,+)}$.

\section{Parametric approximation}
\label{sec:parametric-approx}

In the context of a composite system $\mathcal H_1\otimes \mathcal H_2$ evolving according to a Hamiltonian $\hat H = \hat H_1 + \hat H_2 + \hat H_{12}$, the parametric approximation consists of assuming that system $\mathcal H_1$ evolves according to their free Hamiltonian $\hat H_1$ only, thus neglecting the effect of system $\mathcal H_2$ on it.

\subsection{One branch}
\label{sec:single_branch_param}

Assume the systems initially start at $t=0$ in a separable state
\begin{equation}
    \ket{\Psi}=\ket{\psi}\ket{\phi}.
\end{equation}
The parametric approximation consists of the following two assumptions.
\begin{enumerate}\setlength\itemsep{-.3em}
  \item The state remains separable, that is, for all times, we may write
        \begin{equation}\label{separable-all-times}
            \ket{\Psi(t)}=\ket{\psi(t)}\ket{\phi(t)}.
        \end{equation}
\item  System $\mathcal H_1$ evolves according to its free Hamiltonian:
        \begin{equation}\label{A_free_evol}
            \frac{\dd}{\dd t}\ket{\psi(t)}=-i\hat H_1\ket{\psi(t)}.  
        \end{equation}
\end{enumerate}
Then, explicit differentiation of \eqref{separable-all-times} gives
\begin{equation}
    \frac{\dd}{\dd t}\ket{\Psi(t)}=\frac{\dd}{\dd t}\big(\ket{\psi(t)}\ket{\phi(t)}\big)=\frac{\dd}{\dd t}\big(\ket{\psi(t)}\big)\ket{\phi(t)} + \ket{\psi(t)}\frac{\dd}{\dd t}\ket{\phi(t)},
\end{equation}
while the Schr\"odinger equation gives
\begin{equation}
  \frac{\dd}{\dd t}\ket{\Psi(t)}=-i\hat H\ket{\Psi(t)} = -i(\hat H_1 + \hat H_2 + \hat H_{12})\ket{\psi(t)}\ket{\phi(t)}.
\end{equation}
Equating the right hand sides of these two formulas, using \eqref{A_free_evol}, and finally projecting onto $\ket{\psi(t)}$, we obtain a Schrödinger equation for the state of system $\mathcal H_2$:
\begin{equation}\label{B_driven_evol}
  \frac{\dd}{\dd t}\ket{\phi(t)}=-i\left(\hat H_1+\avg{\hat H_{12}}^{\psi(t)}\right)\ket{\phi(t)}.
\end{equation}
In this equation, the operator $\hat H_{12}$ acting on the composite system $\mathcal H_1\otimes \mathcal H_2$ is replaced by the operator
\begin{equation}
  \avg{\hat H_{12}}^{\psi(t)}=\matrixel{\psi(t)}{\hat H_{12}}{\psi(t)}
\end{equation}
acting only on $\mathcal H_2$. One now can solve for the full evolution of the state of the composite system \eqref{separable-all-times} by first solving the free evolution \eqref{A_free_evol} of $\mathcal H_1$, and then the driven evolution \eqref{B_driven_evol} of $\mathcal H_2$.

\subsection{Several Branches}
\label{sec:multi_branch_derived}

Let us now consider a more general case that allows for entanglement. Let us assume that, that for all times $t\geq0$, the state of the system can be written as
\begin{equation}\label{separable-all-times-i}
  \ket{\Psi(t)}=\sum_ia_i\ket{\psi_i(t)}\ket{\phi_i(t)},
\end{equation}
where, for all $i$,
\begin{equation}\label{A_free_evol_i}
    \frac{\dd}{\dd t}\ket{\psi_i(t)}=-i\hat H_1\ket{\psi_i(t)}.
\end{equation}
Like in the previous case, we take the explicit time derivative of \eqref{separable-all-times-i}, and equate it to the Schr\"odinger equation, using \eqref{A_free_evol_i} to cancel some terms. This time, we obtain
\begin{equation}
  \sum_i a_i\ket{\psi_i(t)}\frac{\dd}{\dd t}\ket{\phi_i(t)}=-i\sum_i a_i\left(\ket{\psi_i(t)}\hat H_2\ket{\phi_i(t)}+\hat H_{12}\ket{\psi_i(t)}\ket{\phi_i(t)}\right).
\end{equation}

Equations \eqref{separable-all-times-i} and \eqref{A_free_evol_i} are analogous to \eqref{separable-all-times} and \eqref{A_free_evol}, respectively. We see that in this case we need two extra assumptions to get a separate Schr\"odinger equation for each $\ket{\phi_i(t)}$. First, we need to assume that the $\ket{\psi_i(t)}$ are orthogonal. Assuming that they are, then, (or simply ignoring their overlap), we obtain
\begin{equation}
        \frac{\dd}{\dd t}\ket{\phi_i(t)}=-i\hat H_2 \ket{\phi_i(t)}-\sum_{j}a_j\matrixel{\psi_i(t)}{\hat H_{12}}{\psi_j(t)}\ket{\phi_j(t)}.
\end{equation}
We then also assume that
\begin{equation}\label{no_cross_evals}
    \matrixel{\psi_i(t)}{\hat H_{12}}{\psi_j(t)} = \delta_{ij}\matrixel{\psi_i(t)}{\hat H_{12}}{\psi_i(t)}.
\end{equation}
Therefore, given the two assumptions \eqref{separable-all-times-i} and \eqref{A_free_evol_i}, the orthonormality of the $\ket{\psi_i}$ and the equation above, we obtain the equation we seek:
\begin{equation}
        \frac{\dd}{\dd t}\ket{\phi_i(t)}=-i\left(\hat H_2 + \avg{\hat H_{12}}^{\psi_i(t)}\right) \ket{\phi_i(t)}.
\end{equation}

In the system we consider in this work, the assumptions of orthogonality and \eqref{no_cross_evals} will be both taken care of by the presence of the control qudits.

\section{Quantum field, quantum controlled sources}
\label{app:quantum_quantum}

We now apply the previous discussion to the case relevant to the main text.

\subsection{Setup}

We may write the Hamiltonian of the full system as
\begin{equation}\label{total_Hamiltonian}
  \hat H(t) = \hat H_A(t) + \hat H_B(t) + \hat H_0 + \hat H_\mathrm{int},
\end{equation}
where $\hat H_A(t)$ acts only on particle $A$ and the first qudit, $\hat H_B(t)$ acts only on particle $B$ and the second qudit, $\hat H_0$ is the free Hamiltonian of the field and $\hat H_\mathrm{int}$ couples both particles to the field.

The time dependence in $\hat H_A(t)$ represents a possible time-dependent driving of the particle $A$ by the first qubit. We assume that
\begin{equation}
  \hat H_A(t) = \sum_{r=1}^d\ketbra rr\otimes \hat H^r_A(t),
  \end{equation}
where $\ket{r}$ are the eigenstates of the qudit in the computational basis. The same holds for for $\hat H_B(t)$.
The free Hamiltonian of the field is the standard
\begin{equation}
  \hat H_0 = \int\dbar^3k\,\omega_k\hat a^\dagger_\k\hat a_\k,
\end{equation}
with $\omega_k=\sqrt{m^2+k^2}$. Finally, the interaction Hamiltonian couples each particle to the field
\begin{equation}
  \hat H_\mathrm{int} = \int\dd^3x\,\hat\phi(\x)\hat\rho(\x),
\end{equation}
where $\hat \rho(\x) = \hat \rho_A(\x) + \hat \rho_B(\x)$ is the sum of two charge density operators, defined as
\begin{equation}\label{charge_density_operator}
  \hat \rho_A(\x) = \int\dd x_A \, \sigma_A(\x-\x_A)\ketbra{\x_A}{\x_A},
\end{equation}
with $\ket{\x_A}$ the position eigenstates of particle $A$, and $\sigma_A(\x)$ a real-valued function with compact support around $\x=\bm 0$ representing the charge distribution of particle $A$; $\hat \rho_B$ is defined similarly.

\subsection{Two parametric approximations}

Write the initial state of the system as
\begin{equation}
   \ket\Psi = \frac1d\sum_{r,s\in\{1,\dots,d\}}\ket{rs}\ket{\Psi^{rs}},
\end{equation}
where $\ket{\Psi^{rs}}$ is a state for the sources and the field. We take the multi-branch parametric approximation of \ref{sec:multi_branch_derived}, with the qudits evolving freely. This represents the qudits acting as controls on the evolution of the particles and field, without suffering a back-reaction.

In practice, we assume that we can write the state at all times as
\begin{equation}
   \ket{\Psi(t)} = \frac1d\sum_{r,s\in\{1,\dots,d\}}\ket{rs}\ket{\Psi^{rs}(t)}.
\end{equation}
This is the combination of both \eqref{separable-all-times-i} and \eqref{A_free_evol_i} (since $\ket{rs}$ has no free Hamiltonian in this case, its free evolution is to stand still).
The different states $\ket{rs}$ of the control qudits are naturally orthonormal, so we are left to verify that \eqref{no_cross_evals} holds, or, in other words that
\begin{equation}
    \matrixel{rs}{\hat H(t)}{r's'} = \delta_{rr'}\delta_{ss'}\matrixel{rs}{\hat H(t)}{r's'}.
\end{equation}
This is indeed the case, since $\hat H_0$ and $\hat H_\mathrm{int}$ act trivially on the control qudits, and the $\hat H_A$ and $\hat H_B$ are block-diagonal on the computational basis. All the assumptions for the multi-branch parametric approximation hold, and we therefore obtain a Schr\"odinger equation for the sources and the field for each value of the control qudits:
\begin{equation}\label{first_param}
  \ket{\Psi^{rs}(t)} = -i\hat H^{rs}(t)\ket{\Psi^{rs}(t)},
\end{equation}
where $\hat H^{rs}(t)=\hat H_A^r(t) + \hat H_B^s(t) + \hat H_0 + \hat H_\mathrm{int}$.

\medskip

We can now solve the $d^2$ equations \eqref{first_param} by taking single branch parametric approximations. We assume that, at all times,
\begin{equation}
  \ket{\Psi^{rs}(t)} = \ket{\psi_A^r(t)}\ket{\psi_B^s(t)}\ket{\psi_\phi^{rs}(t)},
\end{equation}
and that
\begin{equation}\label{psi_matter_prescribed}
  \frac{\dd}{\dd t}\ket{\psi_A^r(t)}=-i\hat H_A^r(t)\ket{\psi_A^r(t)},~~~~~~~~\frac{\dd}{\dd t}\ket{\psi_B^s(t)}=-i\hat H_B^s(t)\ket{\psi_B^s(t)}.
\end{equation}
It then follows, by the arguments in \ref{sec:single_branch_param}, that
\begin{equation}\label{phi_param}
    \frac{\dd}{\dd t}\ket{\phi^{rs}(t)} = -i\left(\hat H_0 + \hat H_\mathrm{int}^{rs}(t)\right)\ket{\psi_\phi^{rs}(t)},
\end{equation}
where
\begin{equation}
  \hat H_\mathrm{int}^{rs}(t) = \matrixel{\psi_A^r(t)\psi_B^s(t)}{\hat H_\mathrm{int}}{\psi_A^r(t)\psi_B^s(t)}.
\end{equation}
The evolution of $\ket{\phi^{rs}(t)}$ is that of a quantum field coupled to a prescribed time-evolving classical source $\rho^{rs}(t,\x)$ given by
\begin{equation}
  \rho^{rs}(t,\x) = 
  \int\dd^3x_A\,\sigma_A(\x-\x_A)|\psi_A^r(t,\x_A)|^2 
  +  \int\dd^3x_B\,\sigma_B(\x-\x_B)|\psi_B^s(t,\x_B)|^2
  \equiv \rho_A^r(t,\x) + \rho_B^s(t,\x),
\end{equation}
where $\psi_A^r(t,\x_A)\equiv \braket{\x_a}{\psi_A^r(t)}$ is the wavefunction of the centre of mass of the source $A$, in position basis, and similarly for $\psi_B^s(t,\x)$.

\subsection{Criterion for field mediation circuit}

Our parametric approximations lead us to a time-dependent state
\begin{equation}
  \ket{\Psi(t)} = \frac1d\sum_{rs}\ket{rs}\ket{\psi_A^r(t)}\ket{\psi_B^s(t)}\ket{\psi_\phi^{rs}(t)},
\end{equation}
where $\ket{\psi_A^r(t)}$, $\ket{\psi_B^s(t)}$, and $\ket{\psi_\phi^{rs}(t)}$ each obey a separate%
\footnote{The evolution of $\ket{\psi_\phi^{rs}(t)}$ explicitly depends on $\ket{\psi_A^r(t),\psi_B^s(t)}$, but since we know the evolution of the latter in advance, we can write the Schr\"odinger equation for $\ket{\psi_\phi^{rs}(t)}$ with a Hamiltonian that does not act on the particles.}
Schr\"odinger equation, given in \eqref{psi_matter_prescribed} and \eqref{phi_param}. Let $\hat U^r_A(t_1,t_2)$, $\hat U^r_B(t_1,t_2)$, and $\hat U^{rs}_\phi(t_1,t_2)$, be the families of unitary operators that implement\footnote{So that $\ket{\phi^{rs}(t_2)} = \hat U^{rs}_{\phi}(t_1,t_2)\ket{\phi^{rs}(t_1)}$, \emph{etc}.} the evolution determined by these equations.
Then, the unitary operator
\begin{equation}
  \hat U(t_1,t_2) = \sum_{r,s}\ketbra{rs}{rs}\otimes\hat U^s_B(t_1,t_2)\otimes\hat U^r_A(t_1,t_2)\otimes\hat U^{rs}_\phi(t_1,t_2),
\end{equation}
implements the evolution of the full system, since
\begin{equation}
\begin{aligned}
     \hat U(t_1,t_2)\ket{\Psi(t_1)}
     = \frac1d\sum_{r,s}\ket{rs}
        \underset{\ket{\psi_B^s(t_2)}}
        {\underbrace{\left(\hat U^s_B(t_1,t_2)\ket{\psi_B^s(t_1)}\right)}}
        \underset{\ket{\psi_A^r(t_2)}}
        {\underbrace{\left(\hat U^r_A(t_1,t_2)\ket{\psi_A^r(t_1)}\right)}}
        \underset{\ket{\psi_\phi^{rs}(t_2)}}
        {\underbrace{\left(\hat U^{rs}_\phi(t_1,t_2)\ket{\psi_\phi^{rs}(t_1)}\right)}}= \ket{\Psi(t_2)}.
\end{aligned}
\end{equation}
This unitary clearly acts on all subsystems, and it is not a given that it can be written is a subsystem local way. However, if it were the case that
\begin{equation}\label{U_splits}
  \hat U^{rs}_\phi(t_1,t_2) = \hat U^{r}_\phi(t_1,t_2)\circ\hat U^{s}_\phi(t_1,t_2),
\end{equation}
then we could write $\hat U(t_1,t_2)$ in subsystem local form:
\begin{equation}
  \hat U(t_1,t_2) = \hat U_{B\phi}(t_1,t_2) \circ \hat U_{A\phi}(t_1,t_2)
\end{equation}
by defining
\begin{equation}
  \hat U_{A\phi}(t_1,t_2) = \sum_{r=1}^d\ketbra{r}{r}\otimes \hat U^r_A(t_1,t_2)\otimes\hat U^{r}_\phi(t_1,t_2),
\end{equation}
and similarly for $\hat U_{B\phi}(t_1,t_2)$.

\medskip

Therefore, to study the subsystem locality of the full evolution $\hat U$, we need to study the behaviour of in-branch field evolution operator $\hat U^{rs}_\phi$. In \ref{app:quantum-classical}, we show that we can write them as
\begin{equation}
  \hat U^{rs}_\phi(t_1,t_2) = e^{\Omega^{rs}(t_1,t_2)}\hat U^{s}_\phi(t_1,t_2)\hat U^{r}_\phi(t_1,t_2)e^{-i\hat H_0(t_2-t_1)}
\end{equation}
(cf. equation \eqref{U-phi-cross-physical-compact}),
 where $\hat H_0$ is the free Hamiltonian of the field and $\Omega^{rs}(t_1,t_2)$ is a pure imaginary number. This is almost of the form \eqref{U_splits}; the main problem is the $e^{\Omega_2^{rs}}$ term.
Therefore the unitary evolution of the full system can be written in  field mediation form---\textit{provided} that the phase $\Omega^{rs}(t_1,t_2)$ vanishes.

In \ref{sec:two_classical_source}, we identify such a sufficient condition: $\Omega^{rs}(t_1,t_2)$ vanishes whenever the support of the expectation values  $\rho_A^r$ and $\rho_B^s$ of the charge densities $\hat \rho_A$ and $\hat \rho_B$ are spacelike separated in all branches.
This is because the phase $\Omega^{rs}(t_1,t_2)$ may be written as
\begin{equation}\label{final_cross_phase}
\begin{aligned}
    \Omega^{rs}(t_1,t_2)=
    &\iint_{t_1}^{t_2}\dd t \dd t'\iint\dd^3 x\dd^3x'\, 
    \rho_A^r(t,\x)\rho_B^s(t',\x')
    \cdot[\hat\phi_I(t,\x),\hat\phi_I(t',\x')] \\
    &+\int_{t_1}^{t_2}\dd t \int_{t_1}^{t}\dd t'\iint\dd^3 x\dd^3x'\, 
    \big(\rho_A^r(t,\x)\rho_B^s(t',\x')+\rho_A^r(t',\x')\rho_B^s(t,\x)\big)
    \cdot[\hat\phi_I(t,\x),\hat\phi_I(t',\x')].
\end{aligned}
\end{equation}
By microcausality, $[\hat\phi_I(t,\x),\hat\phi_I(t',\x')]$ is required to vanish whenever $(t,\x)$ and $(t',\x')$ are spacelike separated. Therefore, if between $t_1$ and $t_2$, $\rho_A^r$ and $\rho_B^s$ are only non-zero in spacelike separated regions, the whole integral vanishes identically.
If this is the case in each $rs$ branch, then full the evolution can be written in field mediation form as
\begin{equation}
  \hat U(t_1,t_2)=\hat U_{B\phi}(t_1,t_2) \circ \hat U_{A\phi}(t_1,t_2)\circ e^{-i\hat H_0 (t_2-t_1)}.
\end{equation}

The charge density expectation values are defined as
\begin{equation}
    \rho_A(t,\x)=\int\dd^3x_A\,\sigma_A(\x-\x_A)|\psi_A^r(t,\x_A)|^2,
\end{equation}
and similarly for $\rho_B$, where $\sigma_A$ is a real-valued function with compact support, representing the charge distribution of the particle. Due to quantum uncertainty, in non-relativistic quantum mechanics, the function $\psi_A^r$ will not have compact support for any finite time, so neither will $\rho_A$. However, one can take an approximation where $|\psi_A|^2$ is infinitesimal outside some region, and ignore its contribution. At that level of approximation, then, $\rho_A$ will have compact support, and one can talk about spacetime separation.

\section{Quantum field, classical sources}
\label{app:quantum-classical}

Let us solve the evolution of a quantum field in the presence of a classical source $\rho(t)$. We may rewrite the Hamiltonian in the Schr\"odinger picture for this system as
\begin{equation}\label{quantum-Hamiltonian}
  \hat H_\phi^\rho(t)= \hat H_0 + \hat H_\mathrm{int}^\rho(t)
\end{equation}
and the field $\hat \phi$ and the momentum $\hat \pi$ are expressed as in terms of creation/annihilation operators as usual.

\subsection{Evolution of the field in the presence of a classical source}
\label{app:quantum-classical-1}

We look for the family of propagators $\hat U^\rho_\phi(t_1,t_2)$ satisfying the Schr\"odinger equation
\begin{equation}
  \frac{\partial}{\partial t_2}\hat U^\rho_\phi(t_1,t_2)=-i \hat H_\phi^\rho(t_2)\hat U_\phi^\rho(t_1,t_2)
\end{equation}
and the group property
\begin{equation}\label{group-property}
  \hat U_\phi^\rho(t_2,t_3)\hat U_\phi^\rho(t_1,t_2)=\hat U_\phi^\rho(t_1,t_3).
\end{equation}
To do so, we introduce the interaction picture\footnote{Note that $t_0$ is the reference time at which Heisenberg, Sch\"odinger, and interaction pictures agree; it could be set to $t_0=0$ without loss of generality.} propagators
\begin{equation}\label{IP-propagator-def}
  \hat U_I^\rho(t_1,t_2)= e^{i\hat H_0(t_2-t_0)}\hat U_\phi^\rho(t_1,t_2)e^{-i\hat H_0(t_1-t_0)},
\end{equation}
which satisfy an analogous group property if and only if the Schr\"odinger operators $\hat U(t_1,t_2)$ do, and satisfy a Schr\"odinger equation:
\begin{equation}
  \frac{\partial }{\partial t_2}\hat U_I^\rho(t_1,t_2) = -i\hat H_I^\rho(t_2)\hat U^\rho_I(t_1,t_2),
\end{equation}
where the operators $\hat H_I^\rho(t)$ are defined as
\begin{equation}
  \hat H_I^\rho(t) = e^{i \hat H_0(t-t_0)}\hat H_\mathrm{int}^\rho(t)e^{-i \hat H_0(t-t_0)}=\int\dd^3x \rho(t,\x) \hat \phi_I(t,\x),
\end{equation}
where $\phi_I(t,\x):=e^{i \hat H_0(t-t_0)}\hat \phi(\x)e^{-i \hat H_0(t-t_0)}$ are the interaction picture field operators. The solution is provided by the Magnus expansion
\begin{equation}
  \hat U_I^\rho(t_1,t_2)=\exp\left(\sum_{n=1}^\infty\hat\Omega_n(t_1,t_2)\right),
\end{equation}
where the explicit expression for the series of operators $\hat\Omega_n(t_1,t_2)$ can be found in \cite{blanes2009magnus}. In our case, thanks to the fact that
\begin{equation}
  [\hat H_I(t),\hat H_I(t')]=\int\dd^3x\dd^3y\,\rho(t,\x)\rho(t',\y)[\hat \phi_I(t,\x),\hat \phi_I(t',\y)]
\end{equation}
is a c-number, we have that $\hat \Omega_n=0$, for all $n\geq3$, and
we have
\begin{equation}\label{UI-magnus}
  \boxed{\hat U^\rho_I(t_1,t_2)=e^{\Omega^\rho_2(t_1,t_2)}e^{\hat\Omega^\rho_1(t_1,t_2)},}
\end{equation}
with
\begin{align}\label{Omega1_def}
\!\!\!\!\!\!\hat \Omega^\rho_1(t_1,t_2)
      &:=-i\int_{t_1}^{t_2}\dd t\,\hat H_I^\rho(t)
      =-i\int_{t_1}^{t_2}\dd t\int\dd^3x\,\rho(t,\x)\hat \phi_I(t,\x),\\
\!\!\!\!\!\!\Omega_2^\rho(t_1,t_2)
      &:=-\frac12\int_{t_1}^{t_2}\dd t\int_{t_1}^{t}\dd t'\,[\hat H_I^\rho(t),\hat H_I^\rho(t')]
      =-\frac12\int_{t_1}^{t_2}\dd t\int_{t_1}^{t}\dd t' \iint\dd^3x\dd^3x'\,\rho(t,\x)\rho(t',\x')[\hat \phi_I(t,\x),\hat \phi_I(t',\x')].
\label{Omega2-def}
\end{align}
Note that $\Omega_2^\rho(t_1,t_2)$ is a c-number.
The operators
\begin{equation}\label{U-given-UI}
  \hat U^\rho_\phi(t_1,t_2)= e^{-i\hat H_0(t_2-t_0)}\hat U_I^\rho(t_1,t_2)e^{i\hat H_0(t_1-t_0)},
\end{equation}
with $\hat U_I^\rho(t_1,t_2)$ given by \eqref{UI-magnus}, give the evolution of the field in the presence of a classical source.

\subsection{Physical interpretation}
\label{sec:physical-interpretation}

We can rewrite the Schr\"odinger picture propagator using \eqref{U-given-UI} and the Magnus expansion as
\begin{equation}\label{nice-U-quantum-classical}
    \boxed{\hat U^\rho_\phi(t_1,t_2)=e^{\Omega^\rho_2(t_1,t_2)}\hat{\mathcal{D}}^\rho{(t_1,t_2)}e^{-i\hat H_0(t_2-t_1)},}
\end{equation}
where we have defined the operator
\begin{equation}\label{D_rho_def}
    \hat{\mathcal{D}}^\rho{(t_1,t_2)}=e^{-i\hat H_0(t_2-t_0)}e^{\hat \Omega^\rho_1(t_1,t_2)}e^{i\hat H_0(t_2-t_0)}=
    \exp\left(e^{-i\hat H_0(t_2-t_0)}\hat \Omega^\rho_1(t_1,t_2)e^{i\hat H_0(t_2-t_0)}\right).
\end{equation}
We will show that $\hat{\mathcal{D}}^\rho{(t_1,t_2)}$ is a displacement operator. Therefore the full evolution $\hat U^\rho_\phi(t_1,t_2)$ consists of the free evolution of the field followed by a displacement operator, up to a global phase that depends only on the classical source.

Note that in classic introductory texts, such as \cite{cohen-tannoudji1989photons,gerry2005introductory} the overall phase factor $e^{\Omega_2(t_1,t_2)}$ is omitted when deriving the evolution of a quantum field coupled to a classical source. This factor is essential in ensuring the correct groupoid property $\hat U^\rho_\phi(t_2,t_3)\hat U^\rho_\phi(t_1,t_2)=\hat U^\rho_\phi(t_1,t_3)$. In the quantum case, this factor is also essential in deriving the correct evolution. However, the phase can be safely neglected when one is interested in a classical source, and evolving only from an initial time to a final time.

To see that $\hat{\mathcal{D}}^\rho{(t_1,t_2)}$ is a displacement operator, we start by writing $\hat\Omega^\rho_1(t_1,t_2)$ in terms of the creation/annihilation operators. Using the property $e^{i\hat H_0T}\hat a_\k e^{-i\hat H_0T}=\hat a_\k e^{-i\omega_k T}$, we may write
\begin{equation}\label{displacement-alpha-rho}
    e^{-i\hat H_0(t_2-t_0)}\hat \Omega^\rho_1(t_1,t_2)e^{i\hat H_0(t_2-t_0)} = -i\int_{t_1}^{t_2}\dd t\int\dd^3x\,\rho(t,\x) \int\frac{\dbar^3k}{\sqrt{2\omega_k}}\left(\hat a_\k e^{i\k\cdot\x} e^{-i\omega_k(t-t_2)}+\hat a_\k^\dagger e^{-i\k\cdot\x} e^{i\omega_k(t-t_2)}\right).
\end{equation}
Now, define
\begin{equation}\label{normal_variable}
    \alpha^\rho(\k;t_1,t_2) =
    -\frac{i}{\sqrt{2\omega_k}}\int_{t_1}^{t_2}\dd t\,e^{-i\omega_k(t_2-t)}\int\dd^3x\,\rho(t,x)e^{-i\k\cdot\x},
\end{equation}
so that
\begin{equation}
    \hat{\mathcal{D}}^\rho{(t_1,t_2)} = \hat D[\alpha^\rho(t_1,t_2)],
\end{equation}
where
\begin{equation}\label{displacement-def}
    \hat D[\alpha] = \exp\left(\int\dbar^3k( \alpha(\k)\hat a_\k^\dagger- \alpha(\k)^*\hat a_\k)\right),
\end{equation}
is a displacement operator.

 The $\alpha^\rho(\k;t_1,t_2)$ have a physical meaning in terms of the classical equations of motions: they are the normal variables of the field perturbation $\phi^\rho(t,\x;t_1)$ caused by the sources from $t_1$ to $t$. In other words, 
 \begin{equation}\label{driven_field}
 	\phi^\rho(t,\x;t_1) = \theta(t-t_1)\int\frac{\dbar^3k}{\sqrt{2\omega_k}}\left(\alpha^\rho(\k;t_1,t) e^{i\k\cdot\x}+\alpha^\rho(\k;t_1,t)^* e^{-i\k\cdot\x} \right)
 \end{equation}
  is the unique solution of the classical Klein-Gordon equation $(\square-m^2)\phi=\rho$ that vanishes for $t\leq t_1$.

We can now characterise the effect of $\hat U^\rho_\phi(t_1,t_2)$ on the creation/annihilation operators, and hence on the field observables. Note that
\begin{equation}
    \hat U^\rho_\phi(t_1,t_2)^\dagger\hat a_\k \hat U^\rho_\phi(t_1,t_2)
        =\hat {\mathcal D}^\rho(t_1,t_2)^\dagger\left(\hat a_\k e^{-i\omega_\k(t_2-t_1)}\right)\hat {\mathcal D}^\rho(t_1,t_2)
        = \hat a_\k e^{-i\omega_\k(t_2-t_1)} + \alpha^\rho(\k;t_1,t_2)
\end{equation}
and that, therefore
\begin{equation}
    \hat U^\rho_\phi(t_1,t_2)^\dagger\hat \phi_I(t_1,\x) \hat U^\rho_\phi(t_1,t_2)
        =\hat \phi_I(t_2,\x)+\phi^\rho(t_2,\x;t_1).
\end{equation}
Therefore, $\hat U^\rho_\phi(t_1,t_2)$ freely evolves the state of the field from $t_1$ to $t_2$ and then displaces it by the classical field contributed by $\rho$ during the interval $[t_1,t_2]$, up to a phase.

\subsection{Evolution of the field in the presence of two classical sources}
\label{sec:two_classical_source}

We now consider the case where $\rho =\rho_A+\rho_B$, and find sufficient conditions to ensure that $\hat U^\rho_
\phi(t_1,t_2)$ split in terms that contain at most one of $\rho_A$ and $\rho_B$ (which will then translate in subsystem-locality in the quantum case).

First, note that $\Omega_2$ splits into three terms, 
\begin{equation}
\Omega^\rho_2(t_1,t_2)=\Omega^{\rho_A}_2(t_1,t_2)+\Omega^{\rho_B}_2(t_1,t_2)+\Omega_2^{\mathrm{cross}}(t_1,t_2;\rho_A,\rho_B),
\end{equation} 
where $\Omega_2^{\rho_A}$ and $\Omega_2^{\rho_B}$ depend only on $\rho_A$ and $\rho_B$, respectively, and the $\Omega^{\mathrm{cross}}$ is
\begin{equation}\label{Omega2_cross}
  \Omega_2^{\mathrm{cross}}(t_1,t_2;\rho_A,\rho_B)=-\frac12\int_{t_1}^{t_2}\dd t\int_{t_1}^{t}\dd t'\iint\dd^3x\dd^3x'\,\big(\rho_A(t,\x)\rho_B(t',\x')+\rho_B(t,\x)\rho_A(t',\x')\big)[\hat \phi_I(t,\x),\hat \phi_I(t',\x')].
\end{equation}
Thanks to the Baker-Campbell-Hausdorff formula, $e^{\hat\Omega^{\rho}_1}$ also splits into three terms
\begin{equation}
  e^{\hat\Omega^{\rho}_1(t_1,t_2)} = e^{\Omega^\mathrm{cross}_1(t_1,t_2;\rho_A,\rho_B)}e^{\hat\Omega^{\rho_B}_1(t_1,t_2)}e^{\hat\Omega^{\rho_A}_1(t_1,t_2)},
\end{equation}
where
\begin{equation}\label{Omega1_cross}
  \!\!\!\!\Omega_1^\mathrm{cross}(t_1,t_2;\rho_A,\rho_B)
    =\frac12[\hat\Omega^{\rho_A}_1(t_1,t_2),\hat\Omega^{\rho_B}_1(t_1,t_2)]
    =-\frac12\iint_{t_1}^{t_2}\dd t\,\dd t' \iint\dd^3x\,\dd^3x'\,\rho_A(t,\x)\rho_B(t',\x')[\phi_I(t,\x),\phi_I(t',\x')].
\end{equation}
Therefore, if we define
\begin{equation}\label{Omega_cross}
  \Omega_\mathrm{cross}(t_1,t_2;\rho_A,\rho_B) = \Omega_1^{\mathrm{cross}}(t_1,t_2;\rho_A,\rho_B) + \Omega_2^{\mathrm{cross}}(t_1,t_2;\rho_A,\rho_B),
\end{equation}
we may write the interaction picture evolution of the field as a product of unitaries that depend on only one of $\rho_A$ or $\rho_B$, up to a phase that depends on both:
\begin{equation}
  \hat U_I^\rho(t_1,t_2)=e^{\Omega_\mathrm{cross}(t_1,t_2;\rho_A,\rho_B)}\hat U^{\rho_B}_I(t_1,t_2) \hat U^{\rho_A}_I(t_1,t_2).
\end{equation}
Repeating the derivation of the previous section, we can put the complete evolution of the field in physical terms
\begin{equation}\label{U-phi-cross-physical}
   \hat U^\rho_\phi(t_1,t_2)=e^{\Omega_\mathrm{cross}(t_1,t_2;\rho_A,\rho_B)}
    \left(e^{\Omega^{\rho_B}_2(t_1,t_2)}\hat{\mathcal{D}}^{\rho_B}_{(t_1,t_2)}\right)
    \left(e^{\Omega^{\rho_A}_2(t_1,t_2)}\hat{\mathcal{D}}^{\rho_A}_{(t_1,t_2)}\right)
    e^{-i\hat H_0(t_2-t_1)}.
\end{equation}
We can write the above in more compact form as
\begin{equation}\label{U-phi-cross-physical-compact}
\boxed{
  \hat U^\rho_\phi(t_1,t_2)=
  e^{\Omega_\mathrm{cross}(t_1,t_2;\rho_A,\rho_B)}
  \hat U^{\rho_B}_\phi(t_1,t_2)\hat U^{\rho_A}_\phi(t_1,t_2)
  e^{-i\hat H_0(t_1,t_2)},
  }
\end{equation}
where $\hat U^{\rho_A}(t_1,t_2)=e^{\Omega^{\rho_A}_2(t_1,t_2)}\hat{\mathcal{D}}^{\rho_A}_{(t_1,t_2)}$, and similarly for $\hat U^{\rho_A}(t_1,t_2)$.

\section{Scalar analog of the GME phases}
\label{sec:scalar-gme}

Consider a coherent state of the field
\begin{equation}
  \ket{f}=\hat D[\alpha_f]\ket0
\end{equation}
peaked around a classical configuration with values
\begin{equation}
  f(\x) = \int\frac{\dbar^3k}{\sqrt{2\omega_k}}\left(\alpha_f(\k)e^{i\k\cdot\x}+\alpha_f(\k)^*e^{-i\k\cdot\x}\right)
\end{equation}
and momentum
\begin{equation}
  \pi_f(\x) = -i\int\dbar^3k\sqrt{\frac{\omega_k}{2}}\left(\alpha_f(\k)e^{i\k\cdot\x}-\alpha_f(\k)^*e^{-i\k\cdot\x}\right)
\end{equation}
under the influence of a classical current $\rho$. We will show that the evolution $\hat U^\rho_\phi(t_1,t_2)$ sends this coherent state to the coherent state peaked around the classical solution $\phi_f^{\rho}$ to the Klein-Gordon equation with source $\rho$ and initial conditions $\smash{(\phi,\dot\phi)=(f,\pi_f)}$:
\begin{equation}
	\hat U^\rho_\phi(t_1,t_2)\ket f = e^{i\theta^\rho_f}\ket{\phi^\rho_f},
\end{equation}
where the phase $\theta^\rho_f$ has a clear physical intepretation as the action of the on-shell trajectory of the field from $\phi^\rho_f(t_1)$ to $\phi^\rho_f(t_2)$:
\begin{equation}
 \theta^\rho_f = S_\phi(\phi^\rho_f;t_1,t_2),
\end{equation}

First, using the definition \eqref{nice-U-quantum-classical} of $\hat U^\rho_\phi(t_1,t_2)$ we have
\begin{equation}
\begin{aligned}
    \hat U^\rho_\phi(t_1,t_2)\ket{f}
    &=e^{\Omega^\rho_2(t_1,t_2)}\hat{\mathcal{D}}^\rho{(t_1,t_2)}e^{-i\hat H_0(t_2-t_1)}\hat D[\alpha_f] e^{i\hat H_0(t_2-t_1)}\ket{0} \\\\
    &=e^{\Omega^\rho_2(t_1,t_2)}\hat{D}[\alpha^\rho(t_1,t_2)] \hat D[\alpha_f e^{-i\omega(t_2-t_1)}]\ket{0},
\end{aligned}
\end{equation}
where we used the fact that $e^{i\hat H_0 t}\ket0=\ket0$ and that $e^{i\hat H_0 t}\hat a_\k e^{-i\hat H_0 t} = \hat a_\k e^{-i\omega_kt}$ to shift the argument of the first displacement operator (defined in \eqref{displacement-def}). This implements the free evolution of $\ket{f}$. Now, merging the two displacement operators, we obtain
\begin{equation}\label{evolve_coherent}
	\hat U^\rho_\phi(t_1,t_2)\ket{f} = e^{i \theta^\rho_f} \ket{\phi_f^{\rho}(t_1,t_2)},
\end{equation}
where,
\begin{equation}
  \ket{\phi_f^{\rho}(t_1,t_2)}=\hat D[\alpha_f e^{-i\omega t}+\alpha^\rho(t_1,t_2)]\ket0,
\end{equation}
and $\phi_f^{\rho}(t_1,t_2)$ is the solution to the Klein Gordon equation with source $\rho$ with initial conditions $(\phi,\dot\phi)=(f,\pi_f)$ at $t_1$.

Let us look at the phase
\begin{equation}
 \theta^\rho_f = -i\big(\Omega^\rho_2(t_1,t_2)+ \varphi^\rho_f(t_1,t_2)\big),
\end{equation}
where $\varphi^\rho_f(t_1,t_2)$ appeared when merging the two displacement operators
using the Baker-Campbell-Hausdorff formula:
\begin{equation}
  \varphi^\rho_f(t_1,t_2) = \frac12\int\dbar^3k\left( \alpha^\rho(\k;t_1,t_2)^* \alpha_f e^{-i\omega(t_2-t_1)}-c.c\right).
\end{equation}
Using the definition \eqref{normal_variable} for $\alpha^\rho$ we  get
\begin{equation}
  \varphi^\rho_f(t_1,t_2) = -\frac i2\int_{t_1}^{t_2}\dd t \int\frac{\dbar^3k}{\sqrt{2\omega_k}} \left(\rho(t,\k)\alpha_f e^{-i\omega(t-t_1)}+c.c\right).
\end{equation}
To evaluate $\Omega_2^\rho$, recall (or verify) that
\begin{equation}
	[\hat \phi_I(t,\x),\hat \phi_I(t',\x') ] = \int\frac{\dbar^3 k}{2\omega_k}\left(e^{i\k\cdot(\x-\x')}e^{-i\omega_k(t-t')}
	-c.c.\right),
\end{equation}
and substitute this in the expression \eqref{Omega2-def} for $\Omega_2^\rho$ to obtain
\begin{equation}
	\Omega_2^\rho(t_1,t_2)=-\frac12\int_{t_1}^{t_2}\dd t\int_{t_1}^t\dd t'\int\frac{\dbar^3k}{2\omega_k}\left(\rho(t,\k)^*\rho(t',\k)e^{-i\omega_k(t-t')}-c.c.\right).
\end{equation}
Using the definition \eqref{normal_variable} for the normal variables of the field generated by $\rho$, we get
\begin{equation}
	\Omega_2^\rho(t_1,t_2)=-\frac i2\int_{t_1}^{t_2}\dd t\int\frac{\dbar^3k}{\sqrt{2\omega_k}}\left(\rho(t,\k)^*\alpha^\rho(\k;t_0,t)+c.c.\right).
\end{equation}

So putting these two together, we get 
\begin{equation}
	\theta^\rho_f  = -\frac 12 \int_{t_1}^{t_2}\dd t\int\dd^3x \,\rho(t,\x)\phi^\rho_f(t, \x; t_1),
\end{equation}
 where we used the relationship between the field's normal variable and its Fourier transform $(\phi(\k)/2 = \alpha(\k)/\sqrt{2\omega_k})$.
 The field sector of the Lagrangian density corresponding to the Hamiltonian $\hat H_0 + \hat H_\mathrm{int}^\rho$ is
 \begin{equation}
 	\mathcal L_\phi = \frac12\phi(\square-m^2)\phi - \rho\phi,
 \end{equation}
and when the field is on-shell and satisfies the field equations, we have
  \begin{equation}
 	\mathcal L_\phi = -\frac12\rho\phi.
 \end{equation}
So we see that the phases generated during the evolution are the values of the on-shell action of the field.

\pagebreak

\twocolumngrid

\bibliography{refs}

\end{document}